\documentclass[10pt, a4paper]{article}

\usepackage{lrec-coling2024} % this is the new style
\usepackage{tipa}
\usepackage{bm}
\usepackage{graphicx}
\usepackage{subcaption}
\usepackage{pifont}
\title{ Exploring Pathological Speech Quality Assessment with  ASR-Powered Wav2Vec2 in Data-Scarce Context}

\name{Tuan Nguyen$^{1}$, Corinne Fredouille$^{1}$, Alain Ghio$^{2}$, Mathieu Balaguer$^{3}$\\  {\bf \large Virginie Woisard$^{3,4,5}$}}

\address{$^{1}$LIA, Avignon Université, Avignon, France\\ $^{2}$Aix-Marseille Univ, LPL, CNRS, Aix-en-Provence, France\\ $^{3}$IRIT, Université de Toulouse, CNRS, Toulouse INP, UT3, Toulouse, France \\ $^{4}$IUC Toulouse, CHU Toulouse, Service ORL de l’Hôpital Larrey, Toulouse, France\\$^{5}$Laboratoire de NeuroPsychoLinguistique, UR 4156, Université de Toulouse, Toulouse, France \\ \{manh-tuan.nguyen, corinne.fredouille\}@univ-avignon.fr\\ alain.ghio@univ-amu.fr, mathieu.balaguer@irit.fr, woisard.v@chu-toulouse.fr }

\abstract{
Automatic speech quality assessment has raised more attention as an alternative or support to traditional perceptual clinical evaluation.
However, most research so far only gains good results on simple tasks such as binary classification, largely due to data scarcity.
To deal with this challenge, current works tend to segment patients' audio files into many samples to  augment the datasets.
Nevertheless, this approach has limitations, as it indirectly relates overall audio scores to individual segments.
This paper introduces a novel approach where the system learns at the audio level instead of segments despite data scarcity.
This paper proposes to use the pre-trained Wav2Vec2 architecture for both SSL, and ASR as feature extractor in speech assessment.
Carried out on the HNC dataset, our ASR-driven approach established a new baseline compared with other approaches, obtaining average $MSE=0.73$ and $MSE=1.15$ for the prediction of intelligibility and severity scores respectively, using only 95 training samples.
It shows that the ASR based Wav2Vec2 model brings the best results and may indicate a strong correlation between ASR and speech quality assessment.
We also measure its ability on variable segment durations and speech content, exploring factors influencing its decision. \\
\newline  }

\begin{document}

\maketitleabstract

\section{Introduction}
\label{sec:intro}

In the past two decades, speech disorder has gained more and more attention from the computer science community.
Not only in helping the patient in daily life task such as speech synthesis, automatic speech recognition,...\citep{intro1}  but also in supporting experts in evaluating the subject's condition \citep{intro2}.
For the assessment of the patient's speech quality, a panel of experts will conduct evaluations based on perceptual information.
The drawback of perceptual methods is that they are really costly, time-consuming and may not be always consistent. 
Therefore, automatic evaluation system based on perceptual information of experts as ground truth has gained attention from the research community.
Automatic speech quality assessment can provide consistent performance compared to human experts.
Hence, this system can encompass a wide range of tasks, from classification (distinguishing between speech disorders and normal speech) to rating aspects such as intelligibility or severity score.\\
Building upon the previous discussion, it is worth noting that the fundamental nature of classification has led to numerous research with promising results.
Conversely, due to the complexity and lack of uniformity in rating scales, regression tasks have often received less attention.
One of the challenges associated with regression is the scarcity of datasets with score rating compared to classification.
Nonetheless, this exacerbates the challenge of developing automated assessment systems, particularly in the context of today's speech technologies, which are primarily data-driven and require a substantial amount of data for processing and generating accurate assessments.\\
This paper focuses more deeply on the domain of regression task, with the goal of exploring assessment scoring using automated systems.
Current research often enriches data by segmenting patients' audio files into multiple samples (excluding non-speech segments) in an effort to expand datasets in order to leverage the use of deep learning.
However, this approach has inherent limitations, as it indirectly associates overall audio scores with individual (smaller) segments, potentially missing crucial context.
Responding to these limitations, we propose an approach in which the system is designed to learn and provide assessment scores on the whole audio rather than focusing on segments.\\
In the context of this paper, two pre-trained Wav2Vec2 models are proposed, each originating from distinct scenarios: one pre-trained via self-supervised learning (SSL) and the other through fine-tuning for automatic speech recognition (ASR).
These pre-trained models will be further fine-tuned to serve as feature extractors for the speech assessment task.
Indeed, we assess the performance of these two architectures for the task of prediction of both intelligibility and severity scores on a speech corpus produced by patients suffering from Head and Neck Cancers. We compare this performance on existing approaches available in the literature, applied on the same corpus.\\
In addition, based on the best pre-trained Wav2Vec2 model we propose, further experiments and analyses were carried out to investigate how the quality of audio files, in terms of duration and therefore content, can impact model performance.

\section{Corpus}
\label{sec:corpus}
The experiments are conducted on two different French speech corpora, C2SI \citep{c2si} and SpeeCOmco \citep{SpeeCOmco,speecomco1}, recorded in the context of Head and Neck Cancers (HNC).
Additionally, a third speech corpus related to the Parkinson's disease, named Aix Hospital Neurology (AHN) corpus \citep{anh1}, is also used to further analyze the generalization capabilities of the system. 
Lastly, Common Voice dataset \citep{commonvoice} is also used with the purpose of creating pre-trained Wav2Vec2 model through ASR tasks.
They will be presented individually in the subsequent sections.
\subsection{C2SI}
C2SI corpus is a set of healthy control (HC) and patients that have been diagnosed with oral cavity or oropharyngeal cancer originating from different tumor locations.
To maintain the stability of their speech impairment throughout the study corpus, patients were required to complete their treatment plan at least 6 months before enrollment and to be in clinical remission.
Both the control and patient groups were instructed to record their speech in variety of tasks including sustaining vowels, picture description, delivering spontaneous speech, pseudo-word or passage reading.\\
In this study, our focus is on the passage reading task.
The participants were asked to read the first paragraph of \textit{La Chèvre de monsieur Seguin}, a short story by Alphonse Daudet.
The audio files were then evaluated on different perceptual criteria by a panel of six expert clinicians.
Within the context of the paper, intelligibility and severity metrics are taken into account.
Each participant is given by a score from 0 to 10 reflecting their level of intelligibility or severity where a score of 0 indicates severe speech disorder or unintelligible speech, while a score of 10 is normal or related to highly intelligible speech.
The final decision (either the score of intelligibility or severity) is considered to be the average of the six scores given by experts.
A set of 105 speakers (84 patients and 21 controls) is used as train and valid set for our automatic system.

\subsection{SpeeCOmco}
\label{sec:SpeeCOmcocorpus}
Speech and communication in oncology (SpeeCOmco) is an additional  corpus dedicated to HNC similar.
It is composed of 27 patients, varying from really severe patients to quite normal ones.
As in the C2SI corpus, these participants were also recorded while performing the same reading task (the first paragraph of \textit{La Chèvre de monsieur Seguin}) among other speech production tasks.
Their speech productions were evaluated by the same panel of expert clinicians using the same metrics.
Therefore, SpeeCOmco is used as a test set in this paper, which has never been exposed to the model during the training phase.

\subsection{Parkinson's disease - AHN}
\label{sec:Parkinson}
15 patients suffering from the Parkinson's disease are involved in this study.
They are part of the larger Aix Hospital Neurology (AHN) corpus, which has 990 dysarthric patients and 160 healthy controls. Most of patients suffer from Parkinson’s disease (601) or Parkinsonian syndromes (98). 
All participants of the AHN corpus were recorded using EVA workstation \citep{evaworkstation} on multiple speech tasks such as sustained vowels, text reading with several speed instructions, spontaneous speech and more.\\
The 15 patients were chosen because they provided diverse reading contexts, useful for further analysis in this paper. 
Indeed, as detailed in \citep{anh1}, they had performed a double task of reading, the first one on the same text reported before (\textit{La Chèvre de monsieur Seguin}) and the second one on an additional French text called \textit{Le Cordonnier}.
This set of 15 patients, exhibiting two different samples of read speech production, aims at assessing the generalization ability of the proposed system and at providing additional analyses considering varying linguistic content.
A panel of eleven expert clinicians listened to these recordings and provided speech quality scores considering severity and intelligibility measurement.
For this corpus, they used a 4-point scale, where 0 represents healthy or intelligible speech respectively, while the opposite represents the most severe condition or the lowest level of intelligibility. 
The mean score for severity is approximately $0.56$, and for intelligibility is $0.3$.
\subsection{Common Voice}
\label{sec:commonvoice}
Common Voice is a multilingual open-source dataset created by Mozilla which is primarily designed for training ASR systems.
It comprises a huge amount of transcribed speech data acquired through the crowdsourcing of reading text.
The French corpus within Common Voice version 6.1 is a set of 375K utterances or approximately 475 hours of audio from 10K French speakers.
This corpus is challenging for ASR system development since it contains different accents, noise and other variation factors.
Because of these reasons, Common Voice is an excellent choice for developing a robust ASR model which is capable of covering multiple aspects of speech signal.
Furthermore, since the audio data in Common Voice is collected from reading activities, it aligns with the same domain of audio that this paper uses for speech assessment.

\section{Proposed Approach}
\label{sec:approach}

Deep learning has recently seen significant developments and achieved impressive results across different domains.
However, when working in the context of pathological speech, one of the most significant challenge is the limitation of data samples, which is crucial for deep learning.
To address this limitation, a potential solution when it comes to data up-sampling is processing speech at the level of individual audio segments.
A single audio recording can be segmented into several smaller segments, where each segment is assigned with the same score as the average score provided by the experts to the entire audio.
Subsequently, the model is trained using these segment samples to generate prediction.
The final decision is determined by averaging the predictions for samples from the same audio.
Additionally, various data augmentation techniques, like speed and tempo distortion, may be applied to the segments.
This method has been applied in certain works \citep{quintas20,robin21} and has shown promising results.\\
In contrast to those promising results, we consider this approach to have a few concerns.
In the first place, presuming the segment (local) score to be identical with overall (global) audio score could lead the system to behave differently than actual expert assessment.
Indeed, each local segment should be assigned a score linked to its context. For instance, a patient might encounter difficulty in pronouncing some particular speech segments like \textipa{m@"sj2 "s@"g\textepsilon{}~("Mist@r s@g\textepsilon)} but not \textipa{nav\textepsilon{} Zam\textepsilon{} y ("nEv@r h\ae d)}.
%MB: Be careful, your ipa conversion is not correct.
Consequently, local scores should be adjusted accordingly, depending on segments and their degrees of production difficulty for the patient.\\
Furthermore, by repeating the same scores to numerous samples could also introduce overfitting issue.
Segment samples may lack of higher-level speech information, such as prosodic prominence, rhythmic group coherence, and temporal dimensions, which are more represented at the entire audio level (depending on the length of segments compared with the entire audio file).
Choosing the right segment duration is also a challenge.
On the other hand, data augmentation techniques, which usually include speech signal modification such as the addition of noise or changes in speech rate, could lead to the loss of important information or an excess of unrealistic data.\\
In order for the automatic assessment system to behave as closely as possible to that of the experts, we propose to train the related model at the audio level without data augmentation. %(Figure~\ref{fig.2}).
By doing this, different important information from speech to non-speech (extra-linguistic features) could be preserved without alteration.
Pre-training technique is proposed here to handle the data scarcity.
\subsection{Wav2Vec2 model}
\label{sec:Wav2Vec2}
Wav2Vec2 \citep{Wav2Vec2facebook} was originally introduced by Facebook as a pre-trained model for SSL task.
The model was learnt on vast amounts of unlabeled audio data, allowing it to extract meaningful representations from audio signals without the need for supervision.
The pre-trained Wav2Vec2 model has demonstrated its effectiveness across various applications due to its ability to learn different dimensions of speech signals \citep{Wav2Vec2layerwise}. 
It can be fine-tuned on small amounts of label data to excel in tasks like speech recognition or speech classification \citep{altowav2vec}.\\
Wav2Vec2 consists of a feature extractor, transformer encoder and quantization block.
The feature extractor, powered by a convolutional network, processes the raw audio into a latent representation. 
Then the transformer encoder captures contextual information and generates continuous embedding from latent space.
Finally, the quantization block quantizes these continuous embeddings, creating an efficient representation for further processing.
In this study, we plan to integrate LeBenchmark Wav2Vec2 large model, as introduced by \citep{LeBenchmark}, which has been pre-trained on French corpus to transfer its knowledge to our assessment system of speech quality by fine-tuning.
This could help the system overcome the data scarcity while providing consistent and reliable results.\\
In the context of feature extraction, as observed in prior research \citep{ sslreview, sslasrpathologies,intro1, vocalwav2vec2} it is evident that Wav2Vec2 pre-trained through self-supervision effectively captures relevant information for the assessment task. 
However, when applied to more complicated assessment task, Wav2Vec2 pre-trained using SSL encounters difficulties in reaching a convergence point \citep{mutliclass}.
On the other hand, one of the well-established methods of speech intelligibility assessment relies on the comparison between the manual transcription of speech signal by expert with the "ground-truth" annotation of the linguistic content produced by speakers.
In this way, different studies \citep{asrintel,asvintel} have shown the potential correlation between intelligibility scores and word error rates computed from the outputs of an ASR system.
With this in mind, we propose the use of Wav2Vec2 fine-tuned for ASR task as a pre-trained model for the assessment system, in comparison to the original Wav2Vec2 pre-trained through self-supervision.
\subsubsection{Wav2Vec2 pre-trained through SSL}
\label{sec:regbasessl}
LeBenchmark introduces to community several pre-trained Wav2Vec2 models via SSL for French language. 
This paper compares the 2 distinct models: \textit{Wav2Vec2-3K-Large}~\footnote{https://huggingface.co/LeBenchmark/Wav2Vec2-FR-3K-large} and \textit{Wav2Vec2-7K-Large}~\footnote{https://huggingface.co/LeBenchmark/Wav2Vec2-FR-7K-large}.
The difference between both models lies on the volume of unlabeled data the model was learnt on.
One model was trained on approximately 3000 hours of healthy speech while the second one was trained on more extensive dataset consisting of 7700 hours of such data.
From now, these two regression models  will be referred as \textbf{3K-SSL} and \textbf{7K-SSL} for ease of reference.
\subsubsection{Wav2Vec2 pre-trained through ASR}
\label{sec:regbaseasr}
Taking the two pre-trained models described in section~\ref{sec:regbasessl}, we fine-tuned them on an ASR downstream task with the Common Voice French dataset.
The 7K model, which serves as an ASR baseline, has been provided by SpeechBrain \citep{SpeechBrain}, an open-source toolkit dedicated to automatic speech processing. 
This model achieved a Word Error Rate (WER) of $9.96\%$ on Common Voice corpus.
For the 3K model, we adopted an end-to-end fine-tuning approach with Connectionist Temporal Classification (CTC) loss function.
After 50 epochs, our ASR model achieved its best WER performance at $13.57\%$.
Finally, we extracted the Wav2Vec2 block from both systems to be placed in the feature extraction in section~\ref{sec:regression}.
Moving forward, we will use the labels \textbf{3K-ASR} and \textbf{7K-ASR} to denote these two models.

\subsection{Speech Assessment Architecture}
\label{sec:regression}
In this section, we look into the system architecture, which combines Wav2Vec2 with additional layers to create an end-to-end solution optimized for regression tasks.
The model architecture can be broken down into three key components:

\begin{enumerate}
\item \textbf{Feature Extractor}: In the initial stage, Wav2Vec2, as detailed in section \ref{sec:Wav2Vec2}, takes on the role of processing raw audio data to derive meaningful representation with dimension of 1024.
Wav2Vec2 will not remain static but will undergo fine-tuning to transfer the knowledge it acquired from its pre-trained task.
The output of Wav2Vec2 from the last layer (layer 24) will be passed to the next intermediate layers.
\item \textbf{Intermediate Layers}: This part includes a pooling layer followed by 2 linear layers.
    \begin{enumerate}
        \item \textit{Pooling layer}: Reduce the temporal dimension from Wav2Vec2 to ensure the consistent shape among data samples using \textit{Statistic Pooling}, which has second-order calculations of standard deviation.
        Other papers \citep{pooling,pooling2} have shown that, by considering both mean and standard deviation, the system can effectively capture not only the information present throughout the whole sequence but also the fluctuations in the data.
        \item \textit{Linear layers}: 2 linear layers with dimension of 1024.
        They are responsible for learning and identifying complicated patterns from the feature extraction stage into a meaningful information connected to final decision.
    \end{enumerate}

\item \textbf{Output Layers}: A simple linear layer with dimension of 1 corresponding to predictive score.
The prediction is evaluated using MSE metric.
\end{enumerate}

\subsection{10-fold validation}
\label{sec:10fold}
A 10-fold validation technique was applied to the training phase due to the data scarcity.
At each fold, $90\%$ of data or approximately 95 speakers were involved in the training process and 10 were set aside as the validation set.
The system was trained for 20 epochs using a small batch size of 1. This training was carried out on a single NVIDIA Tesla A100 with 40GB of VRAM.
This choice of batch size aims to introduce the randomness into the training process \citep{smallbatch2}. %smallbatch,
This randomness was intended to enhance the generalization of model and to prevent the risk of overfitting.
Also due to the size of Wav2Vec2 with more than 300M trainable parameters, using small batch size  helps reduce the computational resource demands.
SpeeCOmco corpus was used as test set similarly for every fold.
The entire process is implemented using SpeechBrain\footnote{https://speechbrain.github.io}.

\section{Results and Discussion}
\label{sec:performance}
% \subsection{Comparing the different feature extractors}
\subsection{Baseline performance}
\label{sec:generalperfromance}
To evaluate the effectiveness of our proposal, we compare our best system with existing works, for which the same dataset, SpeeCOmCo, was involved for both training and testing, permitting comparison.
The first baseline is established by an automatic system using a Shallow Neural Network based on speaker embedding extraction (\textit{x-vectors} or ECAPA-TDNN) \citep{speecomco1}. As reported by the authors, the automatic prediction system provides a best MSE result of  1.75 (RMSE of 1.32 in the paper) for speech intelligibility assessment, and  1.91 (RMSE of 1.38) for speech severity, using reading passage task.
The second baseline relies on a Convolution Neural Network (CNN) based system, trained for a typical French phone classification task to provide a healthy speech representation, and coupled with a Shallow Neural Network for score prediction \citep{sondes,sondes1}. This baseline system, which aims at providing interpretable phonetic knowledge with the prediction score, reaches a best MSE result of 2.97 for speech intelligibility, and 3.05 for speech severity.\\
Remarkably, most of the systems we proposed consistently outperform these existing baseline systems.
With our best model, we achieved between \textbf{58\% to 75\% MSE reduction} compared with the two baselines reported above for intelligibility assessment and between \textbf{40\% to 62\% to  MSE reduction} for severity assessment within the context of SpeeCOmco corpus.
This outstanding performance puts the system at the forefront of the field, highlighting its ability in speech quality assessment.

\subsection{Comparison of feature extractors}
\label{sec:featureperfromance}
The performance of the different regression models, described in sections \ref{sec:regbasessl} and \ref{sec:regbaseasr}, was measured on SpeeCOmco corpus.
The results presented in table~\ref{tab:resultstable} indicate highly promising performance in tasks related to speech intelligibility and speech severity assessment.
Our proposed architectures achieved outstanding results without requiring data augmentation.
Specifically, we obtained an average best MSE at 0.73 for the intelligibility prediction task and 1.15 for the severity prediction task.
Among the four different pre-trained Wav2Vec2 models, it is interesting to note that 3K-SSL model brings the worst performance in term of severity assessment. 
Meanwhile, the 7K-SSL model performs the poorest in the intelligibility prediction task.
It is not entirely clear why these models exhibit distinct behaviors, but our hypothesis is related to approximately 4,700 hours of differing data.
This additional data, sourced from the European Parliament event, may include French non-native speakers who exhibit distinct articulation, accent, or atypical speech patterns.
These differences could cause the slightly lower performance of the 7K-SSL model in intelligibility assessment.
In contrast, they are essential for severity assessment, help the model is more likely to have captured richer speech representations, including vocal bursts, prosody, and other acoustic cues.
Further investigation is required to confirm this hypothesis.\\
\begin{table}[!ht]
\begin{center}
\begin{tabularx}{\columnwidth}{|l|X|X|}
      \hline
      &Intelligibility MSE & Severity MSE\\
      \hline
      3K-SSL& 1.65 $\pm 0.43$ & 2.1 $\pm 0.83$\\
      \hline
      7K-SSL& 1.84 $\pm 0.49$& 1.83 $\pm 0.71$\\
      \hline
      3K-ASR& \textbf{0.73 \bm{$\pm 0.18$}}& \textbf{1.15 \bm{$\pm 0.14$}}\\
      \hline
      7K-ASR& 0.98 $\pm 0.26$ & 1.15 $\pm 0.16$\\
      \hline
\end{tabularx}
\caption{MSE Results on Severity and Intelligibility prediction tasks at the Audio Level according to different pre-trained models (mean and standard deviation considering the 10-fold validation)}
\label{tab:resultstable}
 \end{center}
\end{table}
While comparing feature extractor based on pre-trained SSL with pre-trained ASR, it is surprising that pre-trained ASR extractor outperformed the pre-trained SSL one.
Not only with better average MSE, pre-trained ASR extractor also shows a more consistent performance with a significantly smaller standard deviation.
The system itself is less sensitive to data variability and results in the learning of more robust features.
This finding highlights a concrete connection between the ASR task and speech assessment, potentially shedding new light on future research directions.
This implies the potential for considering the Wav2Vec2 ASR component as a feature extractor.
ASR, being a more specialized task compared to SSL, may offer a more straightforward interpretability, making it a more practical choice for feature extraction.\\
Additionally, it is worth to mention that 3K-ASR model achieves better results compared with 7K-ASR model on both assessment tasks even the 7K-ASR model yields a much better general WER on Common Voice corpus (9.96\%).
This once again emphasizes the argument regarding the bias toward healthy speech in these models.\\
Looking across the tasks, both types of feature extractor perform slightly better and more stable with intelligibility assessment than severity assessment.
However, with ASR based feature extractor, the difference between two tasks narrows, making it the winner.\\
In the following sections, our analysis focuses on the 3K-ASR model, as it obtained the best results and the highest level of stability.
Due to the stability of 3K-ASR model with just 0.18 and 0.14 of standard deviation in both assessment tasks, without losing generality, we can analyze any random fold within the 10-fold model cross-validation for discussion purpose. 
In the following sections, we are discussing the first fold where the model has an MSE of 0.54 for intelligibility and 1.05 for severity in following sections.

\subsection{Prediction evaluation}
\label{sec:analyzeprediction}
A more detail comparison between the prediction scores provided by the 3K-ASR model with the perceptual scores given by the experts will be present in the following sections.
\subsubsection{Intelligibility prediction}
\label{sec:intelpred}
\begin{figure}[!t]
\begin{center}
\includegraphics[width=\columnwidth]{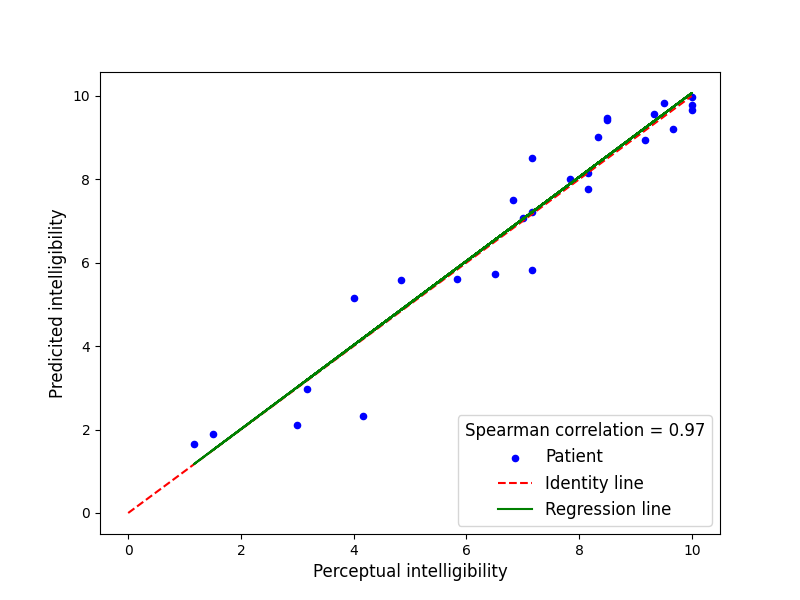} 
\caption{Scatter plot of intelligibility prediction}
\label{fig.intelspear}
\end{center}
\vspace{-0.5cm}
\end{figure}
Figure~\ref{fig.intelspear} illustrates the scatter plot of  intelligibility scores predicted by the automatic system versus the perceptual scores given by the experts.
This visualization provides insight into the relationship between predicted and target scores.\\
On the plot, the red dashed line represents the identity line, where $X=Y$, indicating the perfect match between predicted and target scores.
The green solid line indicates the regression line which shows the best linear relationship between perceptual scores and predictions as determined through regression analysis.\\
The plot clearly shows a remarkably high correlation between the predictions and targets, visualized by the fact that regression and identity lines almost overlap.
This high correlation level is confirmed by Spearman's correlation coefficient of 0.97.
Across all 10-fold validation, the correlation level ranges from 0.94-0.97, with p-values always less than 0.01, highlighting a statistical significance.

\subsubsection{Severity prediction}
\label{sec:sevpred}
\begin{figure}[!t]
\begin{center}
\includegraphics[width=\columnwidth]{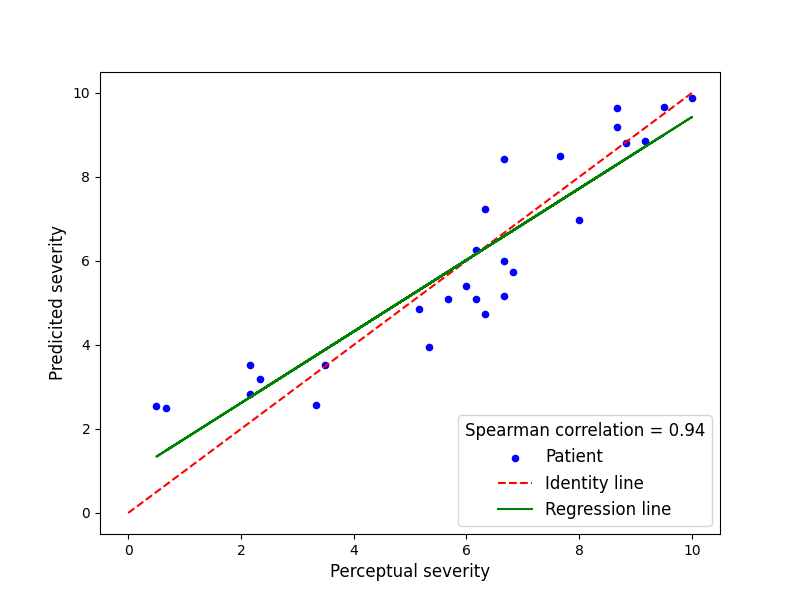} 
\caption{Scatter plot of severity prediction}
\label{fig.sevspear}
\end{center}
\vspace{-0.5cm}
\end{figure}
Similar to section~\ref{sec:intelpred}, figure~\ref{fig.sevspear} depicts the predicted severity scores versus the perceptual scores.
The level of correlation with the severity assessment task is slightly worse than with the intelligibility assessment, as seen in the figure.
The regression line gradually diverges from the identity line, where the lower segment of the regression line tilts upward, and the final segment does the opposite.
It suggests an overestimate for severe patients, as indicated by the lower segment of the plot, and an underestimate for mild patients observed at the point of intersection of the two lines.
\subsection{Generalization and Overfitting}
\subsubsection{Learning curves}
\label{subsec:curve}
\begin{figure}[!t]
\begin{center}
\includegraphics[scale=0.18]{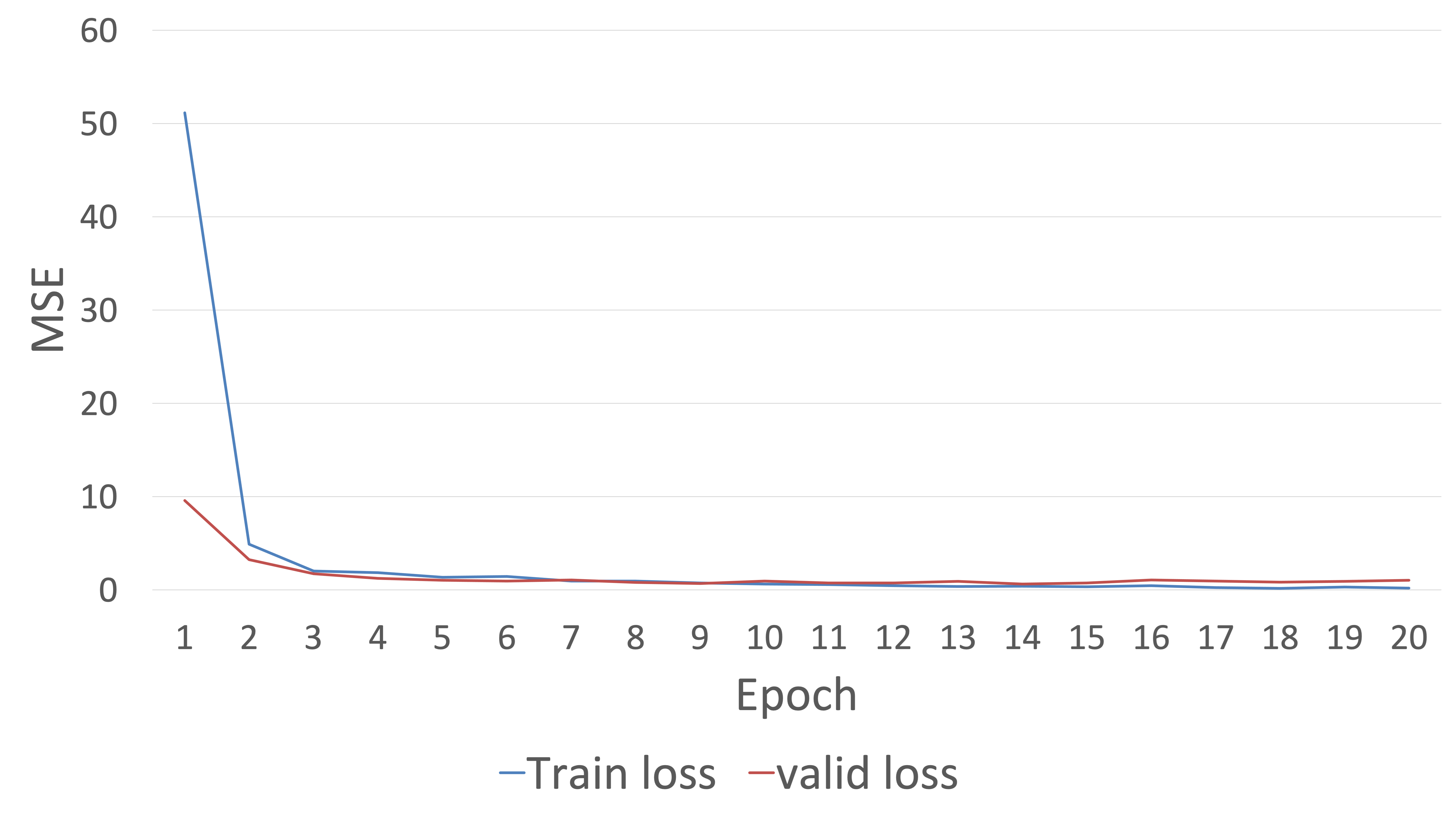} 
\caption{Train and validation loss (MSE) curves from a random fold}
\label{fig.loss}
\end{center}
\vspace{-0.5cm}
\end{figure}

One of the main concerns regarding performance is the possibility of overfitting, particularly when considering the 95 training samples available for each fold and a large model size. However, using 10-fold cross-validation with a fixed test set that the system has never seen before, the model continues to perform well on the test data with high stability.
Furthermore, when examining the loss curve for the intelligibility task, as shown in figure~\ref{fig.loss}, it clearly indicates  that there is no overfitting problem.
At the convergence point, the difference between the training  and validation loss is minimal.
Both loss curves steadily decrease during training until they reach an optimal point.
The same behavior can be observed for the severity performance.
Therefore, it appears that overfitting is not a significant issue, which further strengthens the model's generalization capabilities.
\begin{figure*}[!ht] % Use figure* to span two columns
\centering
\begin{subfigure}[b]{\textwidth} % Adjust the width as needed
    \centering
    \includegraphics[width=\textwidth]{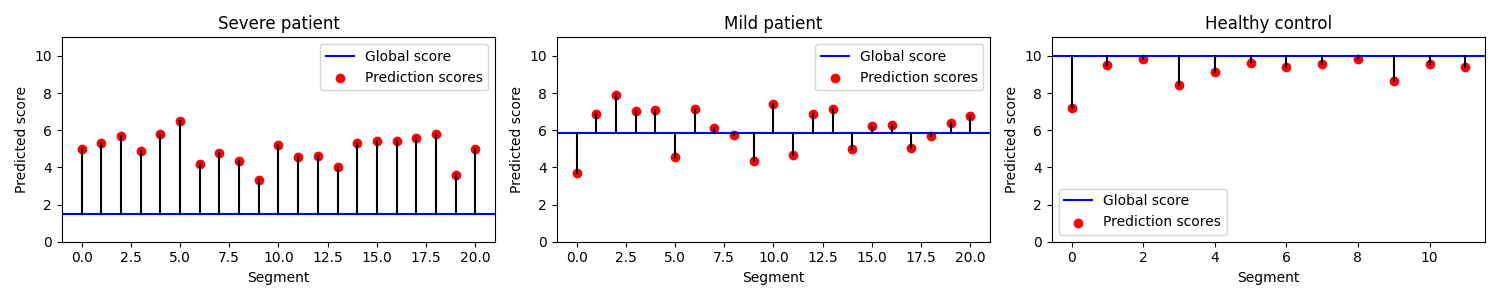}
    \caption{Intelligibility prediction for two-second segment}
    \label{fig:intel_two_sec}
\end{subfigure}

\hfill % Horizontal space between sub-figures
\begin{subfigure}[b]{\textwidth} % Adjust the width as needed
    \centering
    \includegraphics[width=\textwidth]{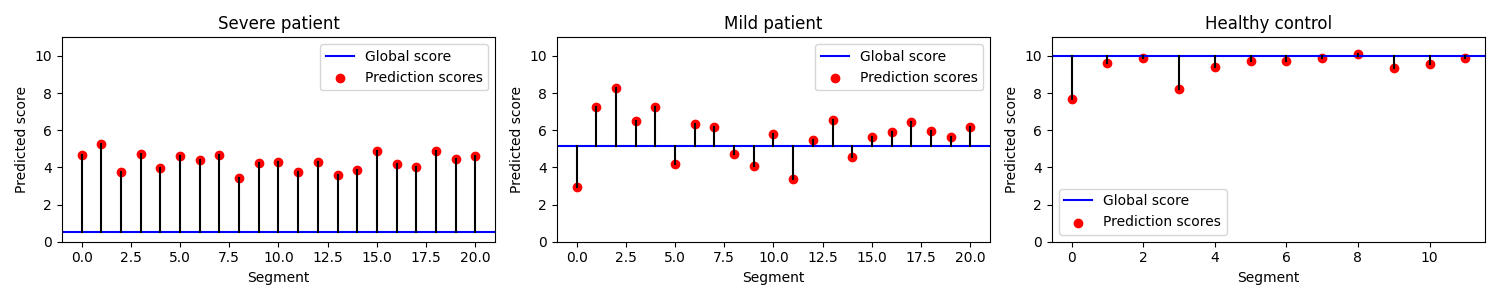}
    \caption{Severity prediction for two-second segment}
    \label{fig:sev_two_sec}
\end{subfigure}
\caption{Model behavior at the segment level}
\label{fig:seg_perfromance}
\end{figure*}

\subsubsection{Cross-domain testing}
Another factor that can reinforce the generalization is the application of Cross-domain testing technique.
To do that, the model is evaluated using data from a different but related domain, specifically, the AHN dataset described in sec~\ref{sec:Parkinson}\\
Normally, patients suffering from Parkinson's disease experience different and relatively mild speech symptoms compared to patients with HNC. This could lead to consider the AHN dataset as a new domain to our system.\\
By evaluating the generalization ability of the 3K-ASR model using the audio \textit{La Chèvre de monsieur Seguin}, we observed a consistent pattern in its predictions for intelligibility and severity assessment.
By converting the scale to a range of 0-10 (which is the scale of the system), we can easily see that the predictions made by 3K-ASR are in line with the perceptual evaluations of experts.
Despite the cross-domain nature, the model consistently shows good performance, achieving an \textit{MSE=0.22} for intelligibility and an \textit{MSE=0.37} for the severity task.
One possible reason why the model achieved better performance with Parkinson's patients compared to SpeeCOmco corpus due to the fact that most of the patients are not severely affected, as indicated by the average score provided in section~\ref{sec:Parkinson}.\\
Based on the findings from the cross-domain corpus and our observations in section \ref{subsec:curve}, we can reinforce our conclusion that overfitting is not a significant concern.

\section{Content impact analysis}
\label{sec:discussion}
This section presents first observations concerning the influences of speech content on the behavior of the speech quality assessment model.
\subsection{Limited content}
As argued in section~\ref{sec:approach}, training the model on the entire audio file ensures that the model can encompass the entire content, context, and various aspects of speech, making the model behavior closer to the global assessment performed by experts.
Furthermore, considering short segments of speech  extracted from the main audio file, the scores may vary depending on the patient's condition.
This section analyzes in detail how the model behaves with shorter test segments  and examines their impact on the model performance.\\
Based on provided arguments, our hypothesis is that the model should provide distinct scores to different segments, depending on the content expressed in those segments.
The average scores across segments are close or relatively close to the global score.\\
For healthy control group, as they do not experience any symptoms, their segment scores should be similar and close to the global score.
This pattern also applies for severe patients, who has extreme difficulty in speaking, resulting in similar low segment scores.
On the other hand, mild patients whose symptoms are not so significant should not experience extreme difficulty in pronouncing all types of phonemes.
Consequently, it is expected that the segment scores for this group vary the most, with both high and low segment scores.\\
To validate this hypothesis, the model generated predictions at a two-second segment level.
Three speakers were carefully selected to ensure that they well represent their respective groups for examination:
\begin{itemize}
    \item \textbf{Severe group}: a patient with perceptual score of 1.5 for intelligibility and 0.5 for severity.
    \item \textbf{Mild group}: a patient with scores of 5.8 for intelligibility and 5.1 for severity.
    \item \textbf{Control group}: a healthy speaker with scores of 10 for both intelligibility and severity.
\end{itemize}
In figure~\ref{fig:seg_perfromance}, the X-axis represents the index of segment along the temporal timeline while the Y-axis indicates model prediction.
The blue line indicates the global score over time while red dots are the predicted scores at the two-second segment level.
The vertical black lines represent the differences between the predicted score and the global score.\\
As expected, figure~\ref{fig:seg_perfromance} demonstrates that the model consistently generated scores on different segments for severe patient and control group.
With the mild group, scores vary more around the reference line.
Nevertheless, when considering the severe patient, the predictions deviate significantly from reference line,  exhibiting an overestimation of the scores by the model.
This indicates that, with limited content information, although the model seems to recognize the consistency in the pathology between segments for severe group, it struggles to make accurate predictions and tends to give an overestimation assessment. 
In a more subtle way, an underestimation of predicted scores can be observed in contrast for the control group.\\
\begin{figure}[!t]
\begin{center}
\includegraphics[width=\columnwidth]{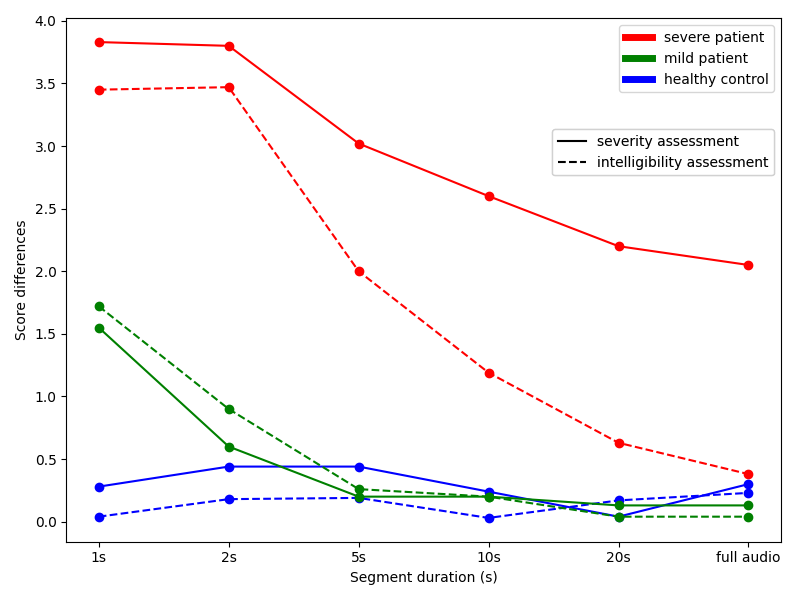} 
\caption{Absolute error variation across different segment durations}
\label{fig.segvariation}
\end{center}
\vspace{-0.5cm}
\end{figure}
To further investigate, we applied the same analysis method to segments of varying duration : one, two, five, ten and twenty seconds with same patients.
The average scores of segments for each patient were compared will the full audio predictions by calculating the absolute error between the targets and these scores.
The results are displayed in figure~\ref{fig.segvariation}.\\
The solid line in the figure indicates the severity assessment task and the dashed line corresponds to the intelligibility assessment.
Patients' speech disorders are represented by different colors: red represents severe patient, green represents mild patient and blue for healthy control.
The graph demonstrates that as the duration increases, the model performance improves, resulting in a reduction in absolute error.
It is logical since increasing duration means that the audio contains more content information and provides the model with more dimension to process.
However, the segment duration does not seem to affect the healthy control group, as the healthy line remains consistent across various durations with good performance.
On the other hand, for severe patient who experiences strong speech disorders, 
there is a close relationship with the amount of contents.
The more contents it has, the better the predictions are.
Same behavior can be observed with the mild group but at a lower level.

\subsection{Different content}
By comparing readings from both texts \textit{La Chèvre de monsieur Seguin} and \textit{Le Cordonnier}, within the AHN corpus, we also observed a high alignment in the predictions made by the automatic system.
Despite the different phonetic contexts of these readings, the system generates high consistent predictions.
Indeed, by applying the Spearman's correlation between the model decisions obtained with the two text readings, we obtained high correlation rates of 0.96 and 0.95 for speech intelligibility and severity assessment respectively, both with a $p$-value of less than 0.01 indicating a statistically significant correlation.
The consistent performance of the model across different contexts indicates that different contents do not affect the final decision.
\section{Conclusion}
This paper proposes a novel approach dedicated to the assessment of speech quality to train model on the entire audio despite the data scarcity. 
To achieve this, we use a regression system with a Wav2Vec2 based model that serves as a feature extractor.
Through experimentation, we find that fine-tuning Wav2Vec2 on ASR yields better results compared to a typical pre-trained Wav2Vec2 SSL when it is fine-tuned for a final regression assessment task.
Only using 95 training samples, we obtained the best result $MSE=0.73$ for intelligibility prediction and $MSE=1.15$ for severity prediction.
Currently, the proposed system outperforms all previous competitors, achieving a significant \textbf{58\% MSE reduction } for intelligibility assessment and a \textbf{41\% MSE reduction} for severity assessment within the context of SpeeCOmco corpus, thereby setting a new performance baseline.
From this, it can be concluded that the ASR pre-trained context is closely related to the speech quality assessment, involving both intelligibility and severity.\\
Moreover, additional analyses showed that the duration of test segments does impact the model decisions. This is particularly true for severe patients; the degree of impact decreases with the patient's speech impairment, as observed for healthy patients who seem less affected.
Regarding now changes in linguistic content (between training and testing), the model does not seem to be significantly affected.
Future work will take a closer look at segment content and, in particular, how certain phonetic contexts might influence the decision of predictive models.
\section{Acknowledgements}
The authors express their heartfelt gratitude to all anonymous reviewers for their insightful comments and suggestions. Additionally, we acknowledge the support of the \textbf{LIAvignon AI Chair}\footnote{https://liavignon.fr} for funding this research work.
\nocite{*}
\section{Bibliographical References}\label{sec:reference}
\bibliographystyle{lrec-coling2024-natbib}
\bibliography{lrec-coling2024-example}
\bibliographystylelanguageresource{lrec-coling2024-natbib}
\let\cleardoublepage\clearpage
\end{document}